\documentstyle[12pt]{article}

\newcommand{\be }{\begin{equation}}
\newcommand{\ee}{\end{equation}}
\newcommand{\ba}{\begin{eqnarray}}
\newcommand{\ea}{\end{eqnarray}}

\newcommand{\non}{\nonumber}

\begin{document}

\title{The  SO(32) Heterotic and Type IIB Membranes}

\author{\\
F. Aldabe${}^{*} {}^{1}$,
and A.L. Larsen${}^{||} {}^{2}$}
\maketitle
\noindent
$^{1}${\em
Theoretical Physics Institute, Department of Physics, \\
University of
Alberta, Edmonton, Canada T6G 2J1.}
\\
\\ $^{2}${\em
Institute of Physics, University of Odense, \\ Campusvej 55, 5230 Odense M,
Denmark.}\\
\begin{abstract}
\baselineskip=1.5em
A two dimensional
anomaly cancellation argument is used to construct
the SO(32)  heterotic  and type IIB membranes.
By imposing different boundary conditions at the two boundaries of a
membrane, we
 shift all of the   two dimensional
anomaly to one of the boundaries.  The topology of these
membranes is that of a 2-dimensional cone propagating in the 11-dimensional
target space.
Dimensional reduction of
these membranes yields the SO(32)  heterotic and type IIB strings.
\end{abstract}
\noindent
\\
\\
$^{*}$Electronic address  faldabe@phys.ualberta.ca\\
$^{||}$Electronic address  all@fysik.ou.dk
\newpage

The current membrane actions could be named the type IIA \cite{howe,
stelle} and $E_8$ \cite{hor, ald}
membranes since their dimensional reduction yields
the type IIA and $E_8\times E_8$ strings, respectively. Recall that the
type IIA membrane
\cite{howe, stelle} has the topology of a torus, while the $E_8$
membrane \cite{hor, ald} has the topology of a cylinder.  To construct the
latter
membrane, one  needs to invoke anomaly cancellation arguments to justify
the coupling
of the membrane to an $E_8$ gauge group at each of its ends.
When the boundary conditions on the membrane are NN (Neumann boundary
conditions at both ends),
 there is a
symmetry between both dynamical boundaries.
Each boundary contributes half of the  two dimensional anomaly present at
the boundaries.

Here we use similar anomaly
arguments to construct the  SO(32) heterotic and type IIB membranes.
By imposing  DN boundary conditions (Dirichlet boundary conditions at one
boundary and
Neumann boundary conditions at the other boundary),
we
 shift all of the   two dimensional
anomaly to the N boundary.  Indeed, the observer at the D boundary will see
no dynamics at all.  Therefore he will be unable to construct any diagram
which contributes to the anomaly.   On the other hand, the observer at the
N boundary sees all the relevant string degrees of freedom.   The observer
at the N boundary  will thus  be
able to construct a
diagram which spoils the two
dimensional coordinate  invariance.

The topology of the  SO(32)  heterotic and type IIB membranes
turns out to be that of a two dimensional cone propagating in the
11-dimensional target space.
That is, one of its boundaries is a string, while the other boundary is a
point in target space.

To fix our notation and conventions, we first give a short review of the
$\kappa$-symmetric action  for the
supermembrane on the  Minkowski background \cite{berg1, berg2}. The action is
\ba
S=-\frac{1}{2}
\int d^3 \zeta \hspace*{-4mm}&\{ & \hspace*{-4mm} \sqrt{-g} (g^{ij}
\Pi^{\mu}_{i} \Pi^{\nu}_{j} \eta_{\mu\nu} -
1) - \epsilon^{ijk} \Pi^\mu_{j} X^\nu_{,k} (
\bar{\theta}\Gamma_{\mu\nu}\theta_{,i}) \non\\
&+& \hspace*{-4mm}\frac{1}{3} \epsilon^{ijk}(\bar\theta\Gamma^{\mu}\theta_{,j})
(\bar\theta\Gamma^{\nu}\theta_{,k})( \bar\theta\Gamma_{\mu\nu} \theta_{,i}) \},
\label{action}
\ea
where,
\be
\Pi^\mu_{i}=X^\mu_{,i}-i\bar\theta\Gamma^{\mu}\theta_{,i}.
\ee
Here $X^\mu=(X^0, X^1,...,X^{10})$ and the world-volume coordinates are
$\zeta^{i}=(\tau,\sigma,z)$. The range of
these coordinates will be specified later.

The  equations of motion corresponding to the action  are
\be
g_{ij}=\Pi^{\mu}_{i} \Pi^{\nu}_{j} \eta_{\mu\nu},
\ee
\be
(\sqrt{-g} g^{ij} \Pi^{\mu}_{j})_{,i}+
\epsilon^{ijk}\Pi^\nu_{i}(\bar{\theta}_{,j}\Gamma^{\mu}\;_{\nu}\theta_{,k})=0,
\ee
\ba
g^{ij}\Pi^\mu_{i}(\Gamma_\mu\theta_{,j})&=&\frac{\epsilon^{ijk}}{2 \sqrt{-g}}\{
(X^\nu_{,k}-\frac{2i}{3}\bar\theta\Gamma^{\nu}\theta_{,k})
(\bar{\theta}\Gamma^{\mu}\;_{\nu}\theta_{,i})(\Gamma_\mu\theta_{,j})\non\\
&-&(\Pi^\mu_{,i}X^\nu_{,k}-\frac{1}{3}(\bar\theta\Gamma^{\mu}\theta_{,i})
(\bar\theta\Gamma^{\nu}\theta_{,k}))(\Gamma_{\mu\nu}\theta_{,j})\}.
\ea
Notice that eq.(5) can be written more compactly, see for instance
Refs.\cite{berg1, berg2}.

The boundary terms, that must vanish to obtain the equations of motion
(3)-(5), are
\be
-\int  d^3 \zeta ({\cal P}^{i}_\mu \;\delta X^\mu+{\cal S}^{i}
\;\delta\theta)_{,i},
\ee
where,
\be
{\cal P}^{i}_\mu=\sqrt{-g} g^{ij} \Pi_{\mu j} +\epsilon^{ijk}(
\bar\theta\Gamma_{\mu\nu} \theta_{,j})
(\Pi^\nu_{,k}+\frac{i}{2}\bar\theta\Gamma^{\nu}\theta_{,k}),
\ee
\ba
{\cal S}^{i}&=&-i\sqrt{-g} g^{ij}
\Pi^{\mu}_{j}(\bar{\theta}\Gamma_\mu)-\frac{i}{2}\epsilon^{ijk}
(X^\nu_{,k}-\frac{2i}{3}\bar\theta\Gamma^{\nu}\theta_{,k})
(\bar{\theta}\Gamma^{\mu}\;_{\nu}\theta_{,j})(\bar{\theta}\Gamma_\mu)\non\\
&-&\frac{1}{2}\epsilon^{ijk}(\Pi^\mu_{j}X^\nu_{,k}-\frac{1}{3}(\bar\theta\Gamma^
{\mu}\theta_{,j})
(\bar\theta\Gamma^{\nu}\theta_{,k}))(\bar{\theta}\Gamma_{\mu\nu}).
\ea
It is also convenient to introduce the following notation for a vector $V^\mu$
\be
V^\pm=\frac{1}{\sqrt{2}}(V^0\pm
V^9),\;\;\;\;\;\;\;\;V^M=(V^1,V^2,...,V^8,V^{10}),
\ee
so that
\be
V^\mu W_\mu=V^M
W^M-V^+W^--V^-W^+,\;\;\;\;\;\;\;\;V^M=V_M,\;\;\;V^+=-V_-,\;\;\;V^-=-V_+
\ee
We follow Refs.\cite{berg1, berg2} and make the "light-cone" gauge choice
\ba
&X^+=p^+\tau,\;\;\;\;\Gamma^+\theta=0&\non\\
&g_{a\tau}=0,\;\;\;a=(\sigma,z)&\\
&g_{\tau\tau}=-{\mbox{det}}(g_{ab})\equiv-{\mbox{det}}(h_{ab})\equiv -h&\non
\ea
Notice that the condition $\Gamma^+\theta=0$ effectively reduces to 16 the
number of components of the spinor
$\theta$
\be
\theta=\pmatrix{S_1\cr S_2 \cr S_1 \cr S_2 \cr},
\ee
where $(S_1, S_2)$ are 8-component spinors (for our conventions for the
gamma matrices, see the appendix).

In the gauge (11), the equations of motion reduce to
\be
g_{ab}=X^M_{,a}X^M_{,b}\equiv h_{ab},
\ee
\be
\ddot{X}^M-(hh^{ab}X^M_{,b})_{,a}+p^+\epsilon^{ab}\bar{\theta}_{,a}\Gamma^{-
M}\theta_{,b}=0,
\ee
\be
\Gamma^-\dot{\theta}+i\epsilon^{ab}\Gamma^{-M}\theta_{,a}X^M_{,b}=0,
\ee
where we introduced also $\epsilon^{ab}=\epsilon^{a\tau b}$. To obtain
these equations of motion, various
properties of the gamma matrices have been used. Our conventions for the
gamma matrices are given in the appendix.

The equations of motion must be supplemented by
the gauge constraints
\be
p^+\Pi^-_{a}=\dot{X}^M
X^M_{,a},\;\;\;\;\;\;\;2p^+{\Pi}^-_{\tau}=\dot{X}^M\dot{X}^M+h,
\ee
which imply, among other things, the relation \cite{duff2}
\be
\epsilon^{ab}\dot{X}^M_{,a}X^M_{,b}+ip^+\epsilon^{ab}\bar{\theta}_{,a}\Gamma
^{-}\theta_{,b}=0.
\ee
These conditions can be used to effectively eliminate three bosonic degrees
of freedom, namely $X^{+}$, $X^{-}$
and one more.

Now consider a boundary of the membrane defined by $z$=const. The
corresponding normal vector is
\be
n_{i}=\frac{1}{\sqrt{h^{zz}}}\pmatrix{0\cr 0 \cr 1 \cr},\;\;\;\;\;\;|n|=1.
\ee
Then the boundary terms (6)-(8) reduce to
\be
-\oint_{B}({\cal P}^{i}_\mu\;\delta X^\mu+{\cal
S}^{i}\;\delta\theta)d\Sigma_{i}=
-\oint_{B}({\cal P}^{z}_\mu\;\delta X^\mu+{\cal
S}^{z}\;\delta\theta)d\Sigma_{z},
\ee
where we used that $d\Sigma_{i}\propto n_{i}$, and the boundary corresponds
to $z=$const. Now,
\ba
{\cal
P}^{z}_M&=&hh^{za}X^M_{,a}-p^+\bar{\theta}\Gamma^{M-}\theta_{,\sigma}\non\\
{\cal P}^{z}_-&=&0\\
{\cal
P}^{z}_+&=&-hh^{za}\Pi^-_{a}-\epsilon^{jzk}\bar{\theta}\Gamma^{M-}\theta_{,j
}X^M_{,k}\non
\ea
such that
\ba
{\cal P}^{z}_\mu\;\delta
X^\mu&=&(hh^{za}X^M_{,a}-p^+\bar{\theta}\Gamma^{M-}\theta_{,\sigma})\delta
X^M\non\\
&-&(hh^{za}\Pi^-_{a}-\epsilon^{zjk}\bar{\theta}\Gamma^{M-}\theta_{,j}X^M_{,k
})\delta X^+.
\ea
On the other hand
\be
{\cal S}^{z}\;\delta\theta=p^+X^M_{,\sigma}\bar{\theta}\Gamma^{M-}\delta\theta.
\ee

We shall be interested in a membrane with two boundaries at $z=z_0$ and
$z=0$, respectively. The other spatial
world-volume coordinate $\sigma$ parametrizes an angular direction,
$\sigma\in[0,2\pi]$.

Consider first the boundary $z=z_0$. We use the notation
$M=(I,10),\;\;I=(1,2,...,8)$, and enforce the
boundary conditions
\be
X^{10}=\mbox{const}.,\;\;\;\;\;\;X^{I}_{,z}=0,\;\;\;\;\;\;S_1= S_2\equiv
S,\;\;\;\;\;\;\;\;\;\;\;\;(z=z_0)
\ee
It is straightforward to show that these conditions in fact kill the
boundary terms (19)-(22).
We shall call this boundary of the membrane the N boundary, since the
conditions on the $X^{I}$
coordinates are of Neumann type. Concerning the conditions on the spinor,
we notice that
 the choice  $S_1= S_2$ determines the chirality
of the surviving fermions at the boundary.  Indeed, for
$S_1=S_2$ we have
\be
(I-\Gamma^{11})\theta=0.
\ee
To ensure that the Neumann boundary conditions are compatible with
supersymmetry, we enforce the additional
boundary condition \cite{bec}
\begin{equation}
S_{,z}=0,\;\;\;\;\;\;\;\;\;\;\;\;(z=z_0)
\end{equation}
 At the boundary $z=z_0$, the target space coordinates $X^\mu$ are
functions of (at
most)
$(\tau,\sigma).$  It means that this boundary of the membrane is a closed
string in target space. We now study
this string in more detail.

Using the gauge choice (11), the action (1) takes the form
\be
S_{lc}=-\frac{1}{2}
\int d^3 \zeta \{ h-\dot{X}^M\dot{X}^M-2ip^+\bar{\theta}\Gamma^-\dot{\theta}+
2p^+\epsilon^{ab}\bar{\theta}\Gamma^{-M}{\theta}_{,a}X^M_{,b}\},
\ee
where we also used (13) to put $h_{ab}$ on-shell. From (23), (25) follows
that the fields have expansions near
$z=z_0$
\ba
X^{10}(\tau,\sigma, z)&=&{\mbox{const}}+(z-z_0)+{\cal O}((z-z_0)^2)\non\\
X^{I}(\tau,\sigma, z)&=&X^{I}(\tau,\sigma)+{\cal O}((z-z_0)^2)\\
\theta(\tau,\sigma, z)&=&\theta(\tau,\sigma)+{\cal O}((z-z_0)^2)\non
\ea
where $\theta(\tau,\sigma)=(S,S,S,S),\;S=S(\tau,\sigma)$.
Notice that we used the remaining freedom mentioned after eq.(17) to fix to
1 the coefficient of the linear term in
$X^{10}$.

The action (26) can be written
\be
S_{lc}=-\frac{1}{2}
\int_{0}^{z_0} dz\; S_{lc}(z),
\ee
where,
\be
S_{lc}(z)\equiv \int d\tau d\sigma \{
h-\dot{X}^M\dot{X}^M-2ip^+\bar{\theta}\Gamma^-\dot{\theta}+
2p^+\epsilon^{ab}\bar{\theta}\Gamma^{-M}{\theta}_{,a}X^M_{,b}\}.
\ee
Using (27), it follows that
\be
S_{lc}(z_0)=\int d\tau d\sigma \{ X^{I}_{,\sigma}
X^{I}_{,\sigma}-{X}^{I}_{,\tau}{X}^{I}_{,\tau}+iS^{T}(S_{,\sigma}+S_{,\tau})\}.
\ee
after a constant redefinition of the spinor $S$.

Thus  we can
interprete the action (30) as the action seen by the observer living at the
boundary $z=z_0$.  This action is
nothing but the spacetime part of the heterotic string action \cite{gross1,
gross2} (i.e. without the SO(32) or
$E_8\times E_8$ current algebra). It must be stressed, that to obtain this
string action, we have
enforced particular boundary conditions on the membrane. However, all the
relevant {\it string} degrees of freedom
are still present.

We now turn to the boundary $z=0$. At this boundary, we take the following
boundary conditions
\be
X^{M}=\mbox{const},\;\;\;\;\;\;\; \;\;\;\;\;\;\;\;\;\;\;
\;\;\;\;\;\;\;\;\;\;\; \;\;\;\;(z=0)
\ee
This is in fact sufficient to kill all the boundary terms (19)-(22) at the
boundary $z=0$.
We shall call this boundary of the membrane the D boundary, since the
conditions on the $X^{M}$
coordinates are of Dirichlet type.
Notice that at this boundary, the target space coordinates are at most
functions of $\tau$. In fact, $X^+$
is proportional to $\tau$, while all the other target space coordinates are
constant. This means that this
boundary of the string is  just a point in target space, that is, a
stationary point with momentum only in the
$X^+$ direction.

As in the previous case, the condition (31) is supplemented by an
additional condition
\begin{equation}
\theta=\mbox{const},\;\;\;\;\;\;\; \;\;\;\;\;\;\;\;\;\;\;
\;\;\;\;\;\;\;\;\;\;\; \;\;\;\;(z=0)
\end{equation}
 Then the fields have expansions near $z=0$
\ba
X^{10}(\tau,\sigma,z)&=&{\mbox{const}}+{\cal O}(z)\non\\
X^{I}(\tau,\sigma,z)&=&{\mbox{const}}+{\cal O}(z)\\
\theta(\tau,\sigma,z)&=&{\mbox{const}}+{\cal O}(z)\non
\ea
It follows that  the action
(29) vanishes identically at $z=0$
\be
S_{lc}(0)=0.
\ee

Thus the boundary conditions (23), (31) describe a membrane whose spatial
topology is a two-cone, that is, one
boundary is a string, while the other is a point in target space. Moreover,
we obtained that the dynamics near
the string boundary ($z=z_0)$
is
the spacetime content  of the heterotic string, while there is no dynamics
at all (except for
the momentum $p^+$)  near the point
boundary ($z=0)$. This concludes the discussion of the boundary conditions.

The action at the boundary $z=0$ has no gravitational anomaly because
it  is not $4k+2$-dimensional at that boundary: the topology of the two-cone
there is just a point particle from a worldvolume point of view.
In addition there is no dynamics there and thus it is impossible
to construct any diagram.  From the target space point of view there is also no
anomaly since the spacetime is 9-dimensional.
This follows from the fact that the topology of
the 11-dimensional targetspace is that of 9-dimensional Minkowski space
times a two-cone, necessary for topological stability of the membranes.

The action at the boundary $z=z_0$ clearly has a two dimensional
gravitational  anomaly.
In order
to cancel such anomaly we must introduce additional massless fields.
The anomaly  can be
cancelled at $z=z_0$
for three different
situations.  Adding an SO(32) or $E_8\times E_8$ current algebra, or adding
additional left moving fermions.  The resulting theories
couple a string to the end of the open membrane.  The respective strings
are $SO(32)$  and $E_8\times E_8$ heterotic strings and type IIB strings.
Notice that the type IIB string is consistent with the boundary conditions.
This follows from the fact that half the worldvolume spinors were
"projected out" of the action after satisfying the boundary conditions.  But
the boundary action (such as the
one needed to obtain the type IIB string) does not
itself need to satisfy any boundary conditions.  Therefore is it acceptable to
have the additional field content needed to obtain a type IIB string at the
boundary.

Thus, at $z=z_0$ we have
\be
S'_{lc}(z_0)=S_{lc}(z_0)+S_{an}(z_0)
\ee
where $S_{an}$ is one of the following actions
\ba
S_{an}(z_0)&=&
\int d\tau d\sigma\{ X^K_{,\sigma}X^K_{,\sigma}-
X^K_{,\tau} X^K_{,\tau}\}\label{t1},\\
S_{an}(z_0)&=&\int d\tau d\sigma
\;i{\tilde{S}}^{T}(\partial_\sigma \tilde{S}+\partial_\tau
\tilde {S})\label{t2} .
\ea
In (\ref{t1}), the bosons $X^K$ will generate the SO(32) or the $E_8\times
E_8$ current algebra required to cancel the gravitational anomaly.
In (\ref{t2}) the fermions $\tilde S$ have the same chirality as the
fermions $S$, and are thus also able to cancel the gravitational anomaly.

The dimensional reduction of these membranes leads to the expected string
theories.  Dimensional reduction in the presence of a boundary has
certain peculiarities. Moreover, in the present case we cannot use the standard
double dimensional reduction ansatz since it is incompatible with our
boundary conditions.
However,  consider the membrane
in the limit that  its world-volume becomes a world-sheet
\be
z\in[0,z_0]\to0\cup z_0\label{limit}
\ee
Then, in that limit, the two boundaries overlap
 and the integral over  $z$ in (28)
 collapses to a sum of two terms
\be
\int_0^{z_0} dz\; S_{lc}(z)\rightarrow \; S_{lc}(0)+\; S_{lc}(z_0).
\ee
The first term vanishes, thus, using also (35), the end-result
in the limit (\ref{limit}) is
\be
\int_0^{z_0} dz\; S_{lc}(z)\rightarrow\; S'_{lc}(z_0)
\ee
which reflects the
dynamics of the  $SO(32)$  and $E_8\times E_8$
heterotic strings and the type IIB string, depending on the type
of action $S_{an}$ chosen to cancel the two dimensional anomaly
(see (\ref{t1})-(\ref{t2})).
\vskip 48pt
\hspace*{-6mm}{\bf Acknowledgements }\\
We thank B. Campbell, R. Myers,  and A. Zelnikov for helpfull
discussions.

\vskip 48pt
\appendix
\section{The Gamma Matrices}
\setcounter{equation}{0}
In this appendix we give our conventions for the gamma matrices.   We
follow closely
the conventions of \cite{green}, however some relabeling of the
coordinates will be required.

The $32\times 32$ gamma matrices  are in the Majorana representation
and are purely imaginary.  They are
\ba
\Gamma^0&=&\tau_2\times I_{16}\non\\
\Gamma^I&=&i \tau_1\times \gamma^I, \;\;\;\;\;\;\;\; I=1,...8\non\\
\Gamma^9&=&i \tau_3\times I_{16}\non\\
\Gamma^{10}&=&i \tau_1\times \gamma^9
\ea
where $\tau_i$ are the Pauli matrices, $I_x$ are $x\times x$
identity matrices and the $16\times 16$ real
matrices $\gamma^I$
satisfy
\be
\{\gamma^{{I}},\gamma^{{J}}\}=2\delta^{{I}{J}},\; \;\;\;\;\;\;\;
 \; {I},{J} =1,...8.
\ee
and
\be
\gamma^9=\prod_{I=1}^{8}\gamma^{{I}}.
\ee
This ensures that
\be
\{\Gamma^\mu,\Gamma^\nu\}=-2\eta^{\mu\nu}.
\ee
We now  construct the $spin(8)$ Clifford algebra.\footnote{
This construction is that presented in Appendix 5.B of Ref.\cite{green}}
The matrices $\gamma^{{I}}$ take the form
\ba
\gamma^{\hat{I}}&=&\pmatrix{0& \tilde{\gamma}^{{\hat{I}}}\cr
 -\tilde{\gamma}^{{\hat{I}}}&0\cr },\  {\hat{I}}=1,...7,\non\\
\gamma^{8}&=&\pmatrix{I_{8}& 0\cr
0&-I_{8}\cr },
\ea
where the $8\times 8$ matrices $\tilde{\gamma}^{{\hat{I}}}$ are
antisymmetric and explicitly given by
\ba
\tilde{\gamma}^1&=&-i \tau_2\times\tau_2\times\tau_2\non\\
\tilde{\gamma}^2&=&i 1_2\times\tau_1\times\tau_2\non\\
\tilde{\gamma}^3&=&i 1_2\times\tau_3\times\tau_2\non\\
\tilde{\gamma}^4&=&i \tau_1\times\tau_2\times1_2\non\\
\tilde{\gamma}^5&=&i \tau_3\times\tau_2\times1_2\non\\
\tilde{\gamma}^6&=&i \tau_2\times1_2\times\tau_1\non\\
\tilde{\gamma}^7&=&i \tau_2\times1_2\times\tau_3
\ea
It follows that  $\gamma^{9}$ is given by
\be
\gamma^{9}=\pmatrix{0&-I_{8}\cr
-I_{8}&0\cr }.
\ee
Furthermore
\be
\Gamma^+=\frac{1}{\sqrt{2}}\pmatrix{i & -i \cr i & -i \cr}\times
I_{16},\;\;\;\;\;\;
\Gamma^-=\frac{1}{\sqrt{2}}\pmatrix{-i & -i \cr i & i \cr}\times I_{16},
\ee
such that
\be
(\Gamma^+)^2=(\Gamma^-)^2=1,\;\;\;\;\;\; \{ \Gamma^+,\Gamma^-\}=2.
\ee
Then it is straightforward to show that the condition $\Gamma^+\theta=0$
leads to
\be
\theta=\pmatrix{S_1\cr S_2 \cr S_1 \cr S_2 \cr}.
\ee
Moreover, it follows that
\ba
&\bar{\theta}\Gamma^\mu\partial\theta=0&,\;\;\;\;\;\;\;\;\;\;\;\;
\mbox{unless}\;\;\mu=-\non\\
&\bar{\theta}\Gamma^{\mu\nu}\partial\theta=0&,\;\;\;\;\;\;\;\;\;\;\;\;
\mbox{unless}\;\;\mu\nu=-M
\ea
where $\bar{\theta}=\theta^\dagger\Gamma_0=\theta^{T}\Gamma_0\;$ ($\theta$
is real). Finally notice that
\be
(\Gamma^\mu)^\dagger=\Gamma^0\Gamma^\mu\Gamma^0,\;\;\;\;\;\;\; \;
\Gamma^{11}=\prod_{\mu=0}^{10}\Gamma^{{\mu}}=i\Gamma^{10}.
\ee

\end{document}